*Research Article*

# Structural Transitions in Sheared Electrically Stabilized Colloidal Crystals

**Joachim Kaldasch,[1] Bernhard Senge,[1] and Jozua Laven[2]**

[1] *Technische Universität Berlin, Fakultät III: Lebensmittelrheologie, Königin-Luise-Straße 22, 14195 Berlin, Germany*
[2] *Eindhoven University of Technology, Laboratory of Materials and Interface Chemistry, P.O. Box 513, 5600 MB Eindhoven, The Netherlands*

Correspondence should be addressed to Joachim Kaldasch; kaldasch@lb.tu-berlin.de

A Landau theory is presented for the structural transition of electrically stabilized colloidal crystals under shear. The model suggests that a structural transition from an ordered layered colloidal crystal into a disordered structure occurs at a critical shear stress. The shear induced structural transition is related to a change of the rheological properties caused by the variation of the microstructure which can be verified by scattering experiments. The theory is used to establish the shape of the flow curves. A good qualitative agreement with experimental results can be achieved, while a scaling relation similar to the elastic scaling is established.

## 1. Introduction

The rheology of suspensions containing small solid particles continues to generate great interest not merely because of its relevance to many industrial processes but also because of the theoretical understanding of many-particle systems. Here we want to focus on shear induced transitions in a subclass of suspensions, those with uniform spherical particles carrying an electric charge. In these suspensions the electrostatic interaction is responsible for the creation of long-range periodic crystal structures in equilibrium. They occur as body-centered-cubic (bcc) colloidal lattices for small particles at low ionic strengths or as face-centered-cubic (fcc) lattices for larger particles and higher ionic strengths [1–3].

A large number of rheological investigations have been carried out on model systems of electrically stabilized colloidal suspensions. Experimental techniques allow varying the particle interaction over a wide range by altering the particle size, the volume fraction, the surface charge, and the electrolyte concentration (e.g., [4–9]). The application of numerical simulations (e.g., [10–13]). The simultaneous investigation of the rheological properties and the microstructure by scattering techniques revealed a connection between the microstructure and the flow properties of concentrated colloidal dispersions (e.g., [4, 14–20]). Hoffman [4] first demonstrated that electrically stabilized colloidal suspensions undergo an order-disorder transition accompanied by shear thickening. This microstructural transition can be understood as a disturbance of the balance between stabilizing forces due to the mutual repulsion of the colloidal particles and hydrodynamic forces in a sheared suspension [21–24].

Experimental evidence of order-disorder transitions in low density colloidal crystals induced by shear perturbations were given first by Ackerson et al. [25]. It was found that under shear hexagonal close packed (hcp) layers arrange parallel to the rheometer walls. They slide over each other and melt after applying a critical shear rate (see for a discussion [26]). Similar results were found by Chen et al. [15] and Chow et al. [17]. They studied dense colloidal crystal and found evidence for the intimate relationship between mechanical properties and changes in the underlying microstructure. The investigation of two dimensional colloidal crystal layers confirms this relationship [27].

Starting point of this paper is the statement that order-disorder transitions are inevitably structural transitions. Structural transitions have successfully been described by Landau type theories [28, 29]. Landau suggested that a structural transition is always associated with a change of the symmetry properties of a system. He introduced an order parameter as a measure of the symmetry properties of the internal structure such that its magnitude switches from zero to a nonzero value at the structural transition. Thermodynamic potentials, generally the free energy, can be regarded to be a function of the order parameter. Near the transition the thermodynamic potential can be represented by a functional Taylor expansion of the order parameter. Structural transitions can be understood as a consequence of the variation of this potential as a function of the order parameter. The great advantage of Landau-type theories is capturing the essential physics of structural transition in a simple mathematical form. However, they are qualitative models with a number of free parameters not known a priori. And as a meanfield theory they give only quantitatively correct results when fluctuations in the system are sufficiently small to be neglected.

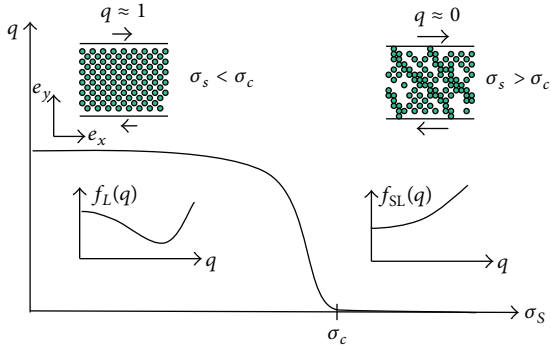

Figure 1: Evolution of the order parameter $q$ at a structural transition as a function of the shear stress $\sigma_S$ around the critical stress $\sigma_C$.

The purpose of the paper is to establish a theoretical model of a shear induced structural transition under the condition that a uniform colloidal crystal structure is present at rest. The paper is organized as follows. In Section 2 the basic equations of the Landau theory of shear induced structural transitions and the relation to the transport parameters, respectively, the structure factor are derived. In Section 3 the theory is applied to experimental investigations followed by a discussion.

## 2. Theory of a Shear Induced Structural Transition

At least three main time scales can be established in a sheared colloidal crystal: the time scale of Brownian motion, the time scale related to the relaxation of the colloidal crystal structure, and the time scale introduced by an applied shear perturbation. The model considers the relaxation time of the colloidal crystal structure and the hydrodynamic perturbation as much longer than Brownian motion. Therefore an undisturbed colloidal crystal is regarded to be in thermal equilibrium.

Suspensions forming colloidal crystals in equilibrium can be treated in a two-medium approximation. The ensemble of spherical, monodisperse, colloidal particles with radius $a$, and number density $\rho(\underline{r})$ is the first medium. They are stabilized, for example, by electrostatic repulsion such that van der Waals attraction can be neglected. The colloidal particles are dispersed in the other continuous medium, the solvent with a viscosity $\eta_S$. Defining the average density of the first medium as $\rho_0$, the colloidal crystal induces periodic variations $\delta\rho(\underline{r})$. The density of a suspension containing a colloidal crystal can therefore be written as

$$\rho(\underline{r}) = \rho_0 + \delta\rho(\underline{r}), \quad (1)$$

where $\delta\rho(\underline{r})$ is a periodic function in the spatial vector $\underline{r}$ due to the colloidal crystal of the form

$$\delta\rho(\underline{r}) = \frac{N}{V} \sum_k \alpha_k e^{i\underline{g_k}\underline{r}} \quad (2)$$

while $N$ is the total number of particles in the sample volume $V$ [29]. Here $k$ runs over the number of the reciprocal lattice vectors $g_k$ of the colloidal crystal and $\alpha_k > 0$ are the corresponding amplitudes of the density waves of the colloidal particle medium. We want to regard the colloidal lattice at rest as consisting of close packed layers arranged parallel to the rheometer walls in a simple shear geometry as displayed in Figure 1 (along $e_x$). The lattice vector can then be separated into a contribution that describes the stacked layers and one that characterizes the internal structure of the hcp-layers. Experimental investigations suggest that when applying a continuous shear perturbation the hcp-layers remain intact but become unstable above a critical stress.

Therefore we introduce an order parameter $q$ of a structural order-disorder transition by

$$q = \sum_k \alpha'_k, \quad (3)$$

where $\alpha'_k$ are the reciprocal lattice amplitudes of the hcp-structure. The value of the order parameter $q$ characterizes therefore the presence of ordered hcp-layers by

$$q \begin{cases} = 0 & \text{disordered structure} \\ > 0 & \text{ordered structure.} \end{cases} \quad (4)$$

That means that when $q$ is unequal to zero close packed layers are present, otherwise the periodic structure vanishes.

Following the standard Landau theory of structural transitions we expand the equilibrium free energy density of a layer $f_L(q)$ as a function of the order parameter [28]:

$$f_L(q) = f_0 + \frac{A'}{2}q^2 + \frac{A''}{3}q^3 + \frac{B'}{4}q^4 + \cdots, \quad (5)$$

where $A'$, $A''$, and $B'$ are functions of the system properties and treated as free parameters. Contributions due to spatial variations of the order parameter are neglected in this derivation. The total free energy of the colloidal crystal at rest is the layer free energy $f_L(q)$ accumulated over all layers.

The parameter $f_0$ represents all contributions to the free energy not related to the order parameter. It can therefore be omitted in the derivation.

We confine our considerations here to the case $|A''| \ll |A'|, |B'|$, such that the third order term can be neglected. The equilibrium value of the order parameter at rest is determined by the minimum of the free energy density with respect to $q$. Hence,

$$q = \begin{cases} 0, & \text{for } A' \geq 0, \; B' > 0, \\ \sqrt{\dfrac{-A'}{B'}}, & \text{for } A' < 0, \; B' > 0. \end{cases} \quad (6)$$

For $A' < 0$, $B' > 0$ the ordered structure is stable. Thus we assume that the layers are present at rest for $A' < 0$.

Applying a continuous simple shear perturbation we assume that the layers remain intact, but the interaction between neighboring layers and with the solvent strains and finally destabilizes the periodic structure of a layer. Above a critical stress $\sigma_C$ the periodic structure will be disrupted associated with an order-disorder transition.

In order to model this shear induced structural transition, we generalize the equilibrium free energy density of a layer given by (5) and introduce the free energy density of a sliding layer $f_{SL}(q, \sigma)$. Since for small shear stresses the interaction between colloidal particles within a layer dominates $f_{SL}(q, \sigma)$, the interaction between neighboring layers and the solvent can be taken into account by adding a small hydrodynamic perturbation $f_H(\sigma, q)$ to (5) such that

$$f_{SL}(q, \sigma) = f_L(q, \sigma) + f_H(\sigma, q) \\ \sim \dfrac{A'}{2} q^2 + \dfrac{B'}{4} q^4 + f_H(\sigma, q). \quad (7)$$

The perturbation $f_H(\sigma, q)$ can be interpreted as a coupling term between the applied shear stress $\sigma$ and the order parameter $q$. Since both parameters are small, we can expand this coupling as a function of the shear stress and the order parameter up to the second order as

$$f_H(\sigma_S, q) = C_0 \sigma_S + C_1 \sigma_S q + \dfrac{C_2}{2} \sigma_S q^2 \ldots, \quad (8)$$

where $C_0$, $C_1$, and $C_2$ are free coupling constants relating the order parameter with the shear stress of the solvent $\sigma_S = \eta_S \dot{\gamma}$, while $\dot{\gamma}$ is the shear rate. The coupling parameters contain the mutual interaction between different layers and the solvent. They are regarded in this approximation as constants. The shear stress is treated as sufficiently small such that higher order terms in $\sigma_S$ can be neglected.

Since the order parameter is not conserved, its evolution under a continuous shear can be approximated by a relaxation dynamics of the form

$$\dfrac{dq}{dt} \sim \dfrac{\partial f_{SL}(q, \sigma_S)}{\partial q}. \quad (9)$$

For a sufficiently fast relaxation of the order parameter the stationary value given by $dq/dt = 0$ becomes with (7) and (8):

$$\dfrac{\partial f_{SL}(q, \sigma)}{\partial q} = A' q + B' q^3 + C_1 \sigma_S + C_2 \sigma_S q = 0. \quad (10)$$

Since we demand that the order parameter disappears in the disordered state, we get for the free parameter $C_1 = 0$. Scaling by $C_2$ we further define

$$A = \dfrac{A'}{C_2}, \qquad B = \dfrac{B'}{C_2}. \quad (11)$$

Setting for convenience

$$A = -B, \quad (12)$$

the order parameter varies between $0 \leq q \leq 1$, while $q = 1$ represents the periodic structure at rest and $q = 0$ corresponds to a disordered state. Solving (10) the order parameter becomes

$$q(\sigma_S) = \begin{cases} 0, & \text{for } \sigma_S \geq -A \\ \sqrt{1 + \dfrac{\sigma_S}{A}}, & \text{for } \sigma_S < -A, \end{cases} \quad (13)$$

while $A < 0$. In difference to equilibrium structural transitions the order parameter becomes a function of the shear perturbation $q = q(\sigma_S)$. Equation (13) describes a shear induced order-disorder transition, which is just a function of the free parameter $A$ and the shear stress $\sigma_S$. The transition takes place at:

$$|A| = \sigma_C = \eta_S \dot{\gamma}_C, \quad (14)$$

where $\dot{\gamma}_C$ is the critical shear rate at the corresponding critical stress $\sigma_C$. The magnitude of the free parameter $A$ characterizes the effective mechanical stability of a periodic layer against a shear perturbation.

The expected evolution of the order parameter $q$ with increasing shear stress is schematically displayed in Figure 1. At equilibrium the free energy of the layers has a minimum at $q \neq 0$ that corresponds to a colloidal crystal. This periodic structure generates Bragg-peaks in a scattering experiment. Applying a continuous simple shear perturbation, the layers slide over each other and the free energy of the layers changes such that its minimum approaches $q = 0$ at the critical shear stress $\sigma_c$. When the layered structure disrupts, the order parameter vanishes accompanied with vanishing Bragg peaks and a variation of the transport properties.

In order to model the rheological properties at such a shear induced transition we take advantage from the shear dependent order parameter and write it as a function of the shear rate $q = q(\dot{\gamma})$, since $\sigma_S$ and the shear rate differ only by the constant solvent viscosity $\eta_S$. The flow properties of the structures on both sides of the transition cannot be derived directly from the Landau theory. Therefore we will take advantage in this paper from phenomenological transport equations. The stationary structures on both sides of the structural transition have specific transport properties. In the limits of fully ordered and disordered structures, the corresponding viscosities $\eta_o(\dot{\gamma})$ and $\eta_d(\dot{\gamma})$ must be retained. For the entire shear rate dependence, we introduce an interpolation equation for the suspension viscosity. Since the shear rate dependent order parameter varies between $0 \leq q(\dot{\gamma}) \leq 1$, the interpolation equation for the viscosity can be formulated as

$$\eta(\dot{\gamma}) = q(\dot{\gamma})\eta_o(\dot{\gamma}) + (1 - q(\dot{\gamma}))\eta_d(\dot{\gamma}). \quad (15)$$

This relation guaranties that the two limiting cases of the stress response of the ordered and disordered structures are retained. In order to capture a general rheological response, both limiting structures are modelled as Bingham plastics. The corresponding phenomenological relations for the ordered, respectively, disordered states are

$$\eta_o(\dot{\gamma}) = \frac{\sigma_o^y}{\dot{\gamma}} + \eta_{io},$$
$$\eta_d(\dot{\gamma}) = \frac{\sigma_d^y}{\dot{\gamma}} + \eta_{id}. \quad (16)$$

Here the constant parameters $\sigma_o^y$ and $\eta_{io}$ are the dynamic yield stress and inherent viscosity of the uniform colloidal crystal and $\sigma_d^y$ and $\eta_{id}$ are the corresponding values of the disordered structure. Note that the disordered structure can be viewed as a colloidal glass with a nonzero yield stress $\sigma_d^y$.

The presented theory allows also characterizing the scattering properties of a monodisperse sheared suspension. The microstructure is determined by the order parameter $q$. Applying scattering techniques to monodisperse suspensions, periodic structures can be identified experimentally by investigating the structure factor $S(\underline{K})$, where $\underline{K}$ is the scattered wave vector. The contribution $\delta S(\underline{K})$ of the periodic arrangement of colloidal particles to the structure factor is determined by

$$\delta S(\underline{K}) \sim \int_V dr^3 \delta\rho(\underline{r}) \exp(-i\underline{K}\,\underline{r}). \quad (17)$$

Substitution of the periodic particle density given by (2) and using (3) leads to

$$\delta S(\underline{K}) \sim q \sum_k \delta\left(\underline{g}'_k(\dot{\gamma}) - \underline{K}\right). \quad (18)$$

The Dirac-delta function in (18) indicates the Bragg-peaks due to the presence of the periodic hcp-layers with reciprocal lattice vectors $\underline{g}'_k$. Choosing the incoming beam perpendicular to the rheometer walls (along $e_y$ in Figure 1) the structure factor exhibits Bragg-peaks that suffer from small periodic variations due to the sliding hcp-layers. The presented Landau theory suggests, however, that when $q$ vanishes, the sheared suspension undergoes an order-disorder transition, in conjunction with the disappearance of the Bragg peaks in (18) and a change of the transport properties described by (15) and (16).

## 3. Results and Discussion

Chen et al. [15] and Chow and Zukoski [17] showed that monodisperse suspensions exhibit a shear induced order-disorder transition by applying simultaneously rheological and small-angle neutron scattering (SANS) techniques. In this section we want to apply the presented Landau theory of shear induced structural transitions to experimental results of Chen et al. [15] and to an extensive rheological study by Fagan and Zukoski [8].

Chen et al. [15] analyzed the flow properties of a suspension of 146 nm diameter charge stabilized latex particles at a volume fraction of $\phi = 0.33$. Simultaneously the microstructure of the sheared suspension was investigated by small-angle neutron scattering. They found that at rest the suspension was ordered, having hexagonally close-packed crystal planes parallel to the rheometer wall. Upon the application of small shear rates the structure at rest is converted into a time-averaged microstructure consisting of a strained crystal in which the hexagonally close-packed layers to the wall remain intact. On further increasing the shear rate Chen et al. [15] found that the suspension changes its structure from a strained crystal structure (SC) into a disordered structure (DS), while they termed the disordered structure as polycrystalline. We prefer the phrase disordered structure in this paper. The transition occurred in conjunction with a change of the transport properties of the sheared suspension at a shear rate of about $\dot{\gamma}_C = 0.1\,\text{s}^{-1}$.

Displayed in Figure 2 is the experimental stress-shear rate dependence of a latex suspension for the case of increasing shear rates as obtained by Chen et al. [15]. The suspension has a pronounced yield stress. With rising shear rate, the suspension initially showed only a slight increase in the stress, followed by a rapid decrease of the shear stress at the critical shear rate $\dot{\gamma}_C = 0.1\,\text{s}^{-1}$. Then the stress increases again with increasing shear rate. The SANS measurements can be summarized by simply separating the stress-shear rate dependence into two branches. Below a shear rate $\dot{\gamma}_C$ a strained colloidal crystal structure (SC) was detected. From $\dot{\gamma} = 0.1\,\text{s}^{-1}$ up to about $\dot{\gamma} = 1\,\text{s}^{-1}$ a disordered structure (DS) has been found.

The presented model allows an interpretation of these rheological measurements as induced by a structural transition of the sheared suspension. The key variable of this instability is the shear rate dependent order parameter $q(\dot{\gamma})$ that is related to the periodic structure of sliding hcp-layers. The presence of the Bragg-peaks in the SANS measurements indicated by (SC) in Figure 2 implies an ordered structure. Since the structural transition can be identified by their disappearance, the vanishing of the Bragg-peaks in the structure factor at the critical shear rate $\dot{\gamma}_C = 0.1\,\text{s}^{-1}$ indicates the order-disorder transition. The shear rate dependent order parameter must be nonzero in the shear rate range $0 \leq \dot{\gamma} < \dot{\gamma}_C$. The order parameter disappears in the disordered structure indicated by (DS) and remains zero for shear rates $\dot{\gamma} > \dot{\gamma}_C$.

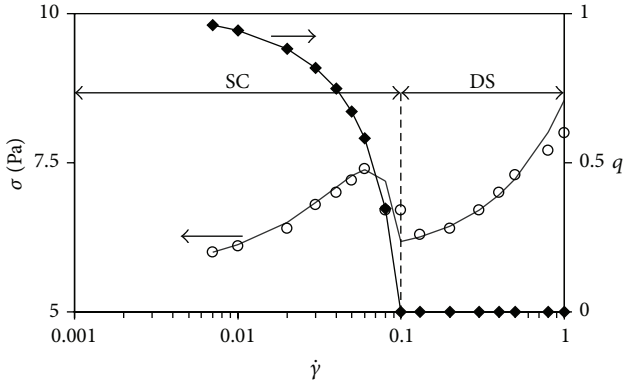

FIGURE 2: Stress (cycles) and estimated order parameter (squares) as a function of the shear rate for latex suspensions with 146 nm diameter particles at a volume fraction of $\phi = 0.33$ (data points taken from [15]). SANS measurements revealed that the flow curve can be separated into two branches: a strained crystal (SC) and a disordered structure (DS) branch. The structural transition occurs at $\dot{\gamma}_C = 0.1\,\text{s}^{-1}$.

Also displayed in Figure 2 is the experimental shear stress as a function of the shear rate (circles). The rheological properties can be obtained by fitting the data with (15) (solid line). A good agreement with the experimental results can be obtained by applying (16) with the fit parameters $\sigma_o^y = 5.7$ Pa, $\eta_{io} = 43.8$ Pa·s for the ordered structure and $\sigma_d^y = 5.9$ Pa, $\eta_{id} = 2.6$ Pa·s for the disordered structure, while $A = -9 \ast 10^{-5}$ Pa. The corresponding evolution of the order parameter determined by (13) is also displayed in Figure 2 (squares).

Increasing the shear rate even further Chen et al. [15] discovered a reappearance of the Bragg-peaks. Chow and Zukoski [17] found a similar result. At high shear rates, however, hydrodynamic forces dominate the dynamics. This shear rate range is therefore out of the scope of the presented model.

We want to apply the theory to another rheological study on sheared colloidal crystals carried out by Fagan and Zukoski [8] on aqueous, monodisperse suspensions of 310 nm silica particles. The experimental results that are displayed in Figure 3 performed at different volume fractions, while we omitted the lowest volume fraction since no structural transition was noticed. The stress-shear rate dependence consists of two branches, separated by an inflection point. This experimental result differs from that of the latex suspension of Chen et al. [15], where two branches are separated by a jump in the shear stress. The modification may be caused by the fact that the particles in Figure 2 are smaller than those in Figure 3. Therefore particles in Figure 3 relax much slower to close packed layers such that they are disturbed. As a consequence the transition between ordered and disorder structures is much less pronounced.

The inflection point indicates in this model the order-disorder transition, which occurs according to (14) at a critical shear stress $\sigma_C = |A|$. The solid lines in Figure 3 are obtained by (15) and (16) with the fit parameters summarized in Figure 4 for the corresponding volume fractions $\phi$. The stability of the periodic structure is characterized by the magnitude of the stress parameter $\sigma_C$. The increase of the magnitude $|A|$ with $\phi$ can be expected, since the interaction between the repulsive colloidal particles increases with increasing volume fraction.

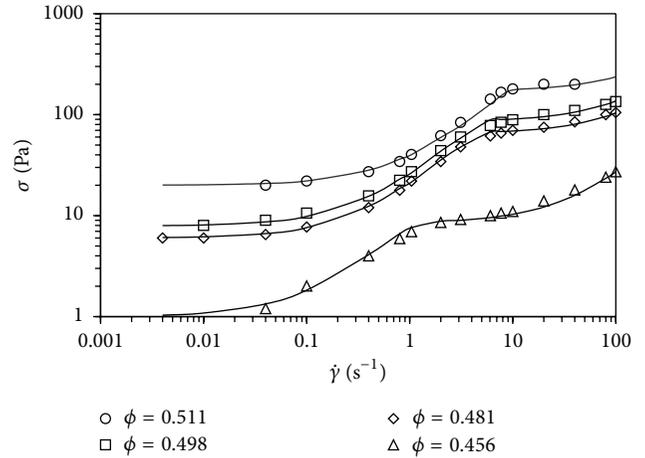

FIGURE 3: Stress as a function of the shear rate for suspensions at four volume fractions of 310 nm diameter silica particles dialyzed to equilibrium against $10^{-3}$ M KCl from Fagan and Zukoski [8]. The solid lines are a fit with the free parameters given in Figure 4.

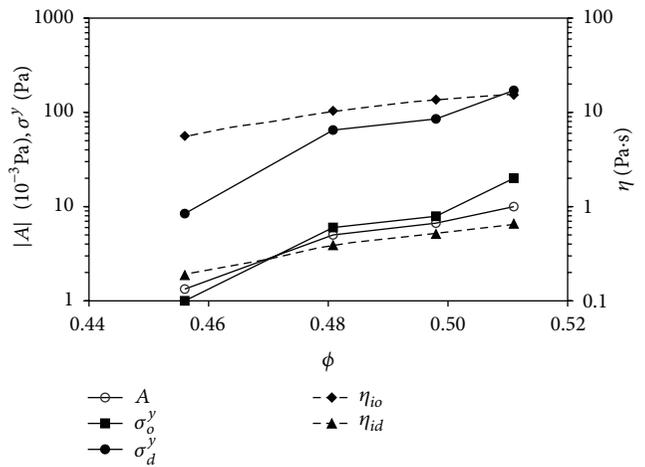

FIGURE 4: The free parameter $|A|$, the inherent viscosities of the ordered and disordered structures $\eta_{io}$, $\eta_{id}$, and the corresponding yield stresses $\sigma_o^y$, $\sigma_d^y$ as a function of the volume fraction $\phi$. The data were obtained from the fit of the data points in Figure 3.

The presented model suggests that the first branch corresponds to a layered colloidal crystal structure with a yield stress $\sigma_o^y$ and inherent viscosity $\eta_{io}$. The second branch represents the disordered structure with constant $\sigma_d^y$ and inherent viscosity $\eta_{id}$. The fit parameters shown in Figure 4 show that the dynamic yield stresses and viscosities of both, the ordered and the disordered structures, increase with increasing volume fraction. The experimentally obtained data exhibit a nearly exponential increase of both, the inherent viscosity and the yield stress with increasing volume fractions. However, the elastic contributions, given by the dynamic

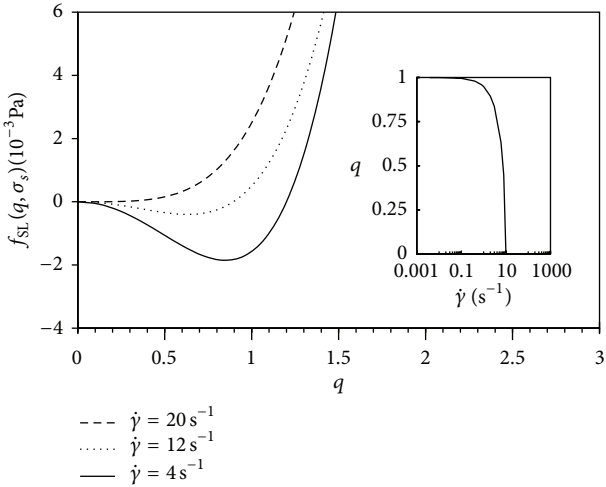

FIGURE 5: The generalized free energy $f_{SL}(q, \sigma_S)$ of hcp-layers in the suspension with $\phi = 0.511$ used in Figure 3, as a function of the order parameter $q$ for different shear rates. The average value of the order parameter corresponds to the minimum of the generalized thermodynamic potential. The insert shows the dependence of the order parameter on the shear rate for this suspension. The microstructural transition occurs when the order parameter vanishes.

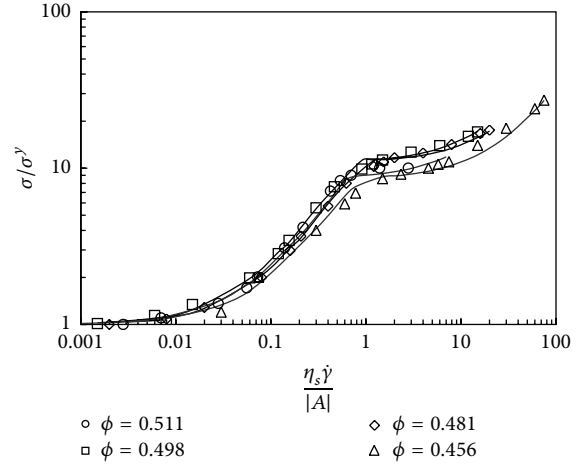

FIGURE 6: Reduced stress as a function of the reduced shear rate, for the 310 nm suspensions in Figure 3, at different volume fractions $\phi$. The flow curves of the volume fractions collapse onto a single master curve, when the shear stress is scaled by the yield stress of the ordered phase $\sigma_o^y$ and the shear rate $\dot{\gamma}$ is replaced by $\eta_S \dot{\gamma}/|A|$. This scaling procedure is similar to the elastic scaling as established by Fagan and Zukoski [8].

yield stresses, increase faster with the volume fraction than the viscosities.

The shear rate dependence of the Landau-potential $f_{SL}(q, \dot{\gamma})$ as suggested by (7) and (8) is depicted in Figure 5 for the volume fraction $\phi = 0.511$. Note that the minimum of $f_{SL}(q, \dot{\gamma})$ corresponds to the time averaged equilibrium value of the order parameter, displayed in the insert of Figure 5 as a function of $\dot{\gamma}$. At $\dot{\gamma}_C = 10\,\text{s}^{-1}$ the order-disorder transition occurs.

Scaling the experimentally obtained shear stresses by the elastic modulus of the crystal structure $G_o$ and replacing the shear rate $\dot{\gamma}$ by $\eta_S \dot{\gamma}/G_o$, Fagan and Zukoski [8] found that the measured flow curves of different volume fractions collapse onto a single master curve. They denoted this scaling procedure as elastic scaling. The scaling relation for the inflection point has therefore the form

$$\text{const.} = \frac{\eta_S \dot{\gamma}_C}{G_o}. \tag{19}$$

Note that this experimental result is in agreement with (14) when we suggest that the free parameter $A$ characterizing the stability of the colloidal crystal is proportional to the elastic shear modulus $|A| \sim G_o$. The almost identical $\phi$-dependence of the dynamic yield stress of the ordered phase and the stress parameter $A$ (see Figure 4) confirms that $\sigma_o^y$ will be roughly proportional to the elastic modulus $G_o$. We can therefore expect that the flow curves for different volume fractions collapse onto a single master curve, when the shear stress is scaled by the yield stress of the ordered phase $\sigma_o^y$ and the shear rate $\dot{\gamma}$ is replaced by $\eta_S \dot{\gamma}/|A|$. That this is indeed the case is shown in Figure 6.

The presented Landau theory of shear induced structural transition suggests the following conclusions.

(1) Sheared, electrically-stabilized suspensions may arrange in layers of close packed colloidal particles parallel to the rheometer walls under a simple shear perturbation. These colloidal crystals can be perturbed and suffer from a shear induced structural transition at a critical shear stress. This instability can be investigated by scattering (e.g., SANS) experiments. The structural transition can be described by a shear dependent order parameter associated to the periodic structure of the hcp-layers and can be identified by the disappearance of the corresponding Bragg peaks in simultaneous scattering experiments. Note that the presented theory is confined to monodisperse colloidal particles.

(2) As a consequence of the shear induced structural transition, the transport properties alter at the instability. Treating the rheological properties of the limiting ordered and disordered structures as Bingham fluids, the shear dependent mechanical properties near the transition can be described with an interpolation relation containing the order parameter as the key variable.

(3) The critical shear stress is governed by the scaling relation equation (14), which is very similar to the experimentally discovered elastic scaling relation equation (19) [8]. Applying the scaling relation the flow dependences of colloidal suspensions at different volume fractions collapse onto a single master curve.

Since the critical stress is related to the stability of the colloidal crystal, (14) suggests that the transition shear rate shifts to higher values when the electrostatic repulsion is increased.

Although the presented model is only of qualitative nature, a good agreement with rheological and scattering data can be achieved. However, in order to understand the impact of particle and solution properties on structural transitions further investigations are indispensable.

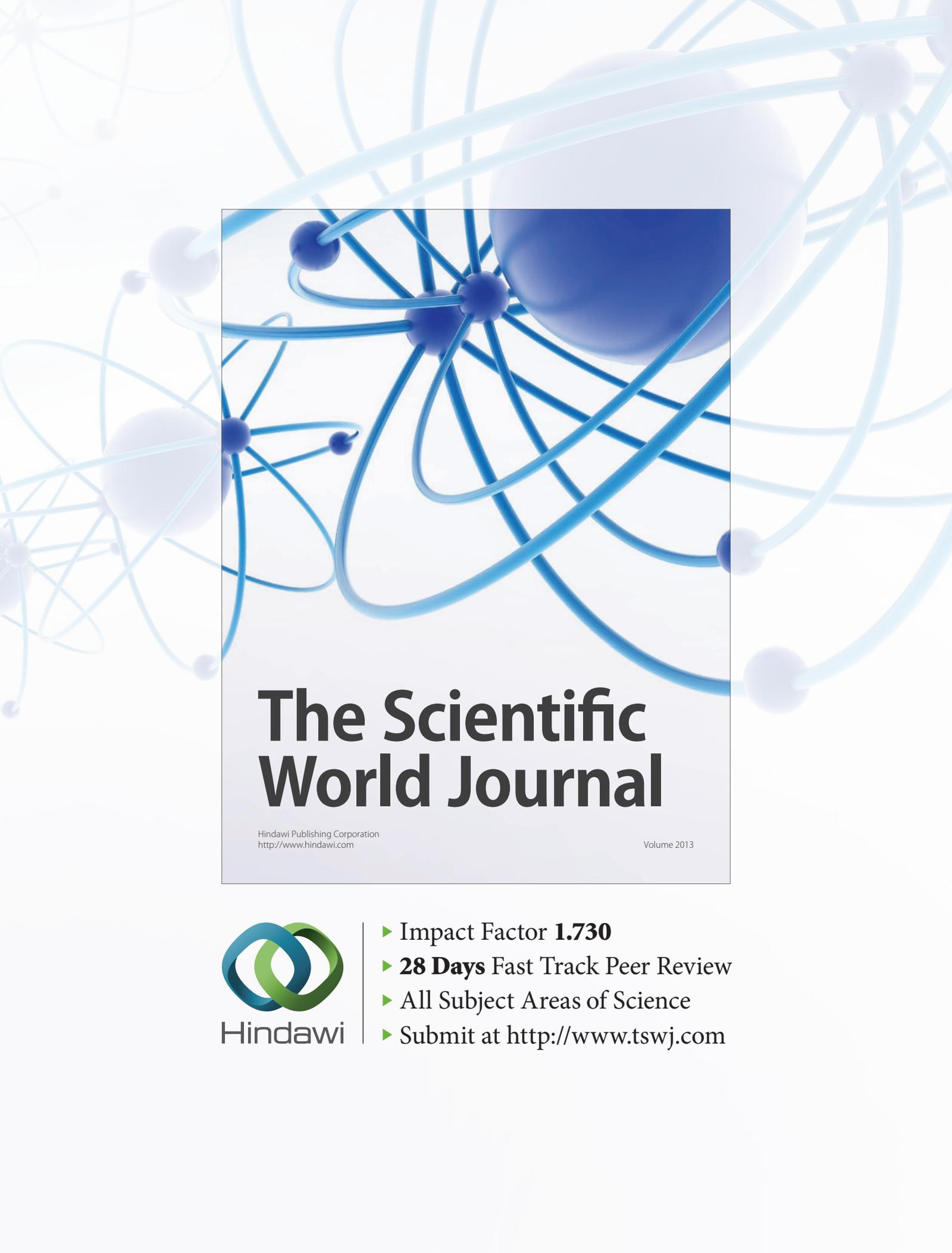